\documentclass[aps,reprint,superscriptaddress]{revtex4-1}
\usepackage{upgreek}
\usepackage{amsmath,amssymb,amsfonts}
\usepackage{appendix}
\usepackage{bm}
\usepackage{color}
\usepackage[normalem]{ulem} 

\usepackage{textcomp}

\usepackage{graphicx}
\usepackage{indentfirst}
\usepackage{psfrag}
\usepackage{epsfig}

\usepackage{color} 

\newcommand{\bra}[1]{\ensuremath{\left\langle#1\right|}}
\newcommand{\ket}[1]{\ensuremath{\left|#1\right\rangle}}

\newcommand{\Veff}{\ensuremath{V_\text{eff}}}
\newcommand{\Veffa}{\ensuremath{V_{\text{eff},\alpha}}}

\newcommand*{\etensor}{\overline{\overline{\varepsilon}}}
\newcommand*{\ctensor}{\tilde{c}}

\begin{document}

\title{Acoustic diamond resonators with ultra-small mode volumes}

	\author{Miko\l{}aj K. Schmidt}
	\email{mikolaj.schmidt@mq.edu.au}
	\affiliation{Macquarie University Research Centre in  Quantum Engineering (MQCQE), MQ Photonics Research Centre, Department of Physics and Astronomy, Macquarie University, NSW 2109, Australia.}
	\author{Christopher G. Poulton}
	\affiliation{School of Mathematical and Physical Sciences, University of Technology Sydney, NSW 2007, Australia.}
	\author{Michael J. Steel}
	\affiliation{Macquarie University Research Centre in  Quantum Engineering (MQCQE), MQ Photonics Research Centre, Department of Physics and Astronomy, Macquarie University, NSW 2109, Australia.}
	
\begin{abstract}
    Quantum acoustodynamics (QAD) is a rapidly developing field of research, offering possibilities to realize and study macroscopic quantum-mechanical systems in a new range of frequencies, and implement transducers and new types of memories for hybrid quantum devices. Here we propose a novel design for a versatile diamond QAD cavity operating at GHz frequencies, exhibiting effective mode volumes of about $10^{-4}\lambda^3$. Our phononic crystal waveguide cavity implements a non-resonant analogue of the optical lightning-rod effect to localize the energy of an acoustic mode into a deeply-subwavelength volume. We demonstrate that this confinement can readily enhance the orbit-strain interaction with embedded nitrogen-vacancy (NV) centres towards the high-cooperativity regime, and enable efficient resonant cooling of the acoustic vibrations towards the ground state using a single NV. This architecture can be readily translated towards setup with multiple cavities in one- or two-dimensional phononic crystals, and the underlying non-resonant localization mechanism will pave the way to  further enhance optoacoustic coupling in phoxonic crystal cavities.

\end{abstract}
\maketitle

\section{Introduction}

New developments in the field of GHz quantum acoustics are closely mirroring those reported in integrated cavity and waveguide quantum electrodynamics. The acoustic toolbox now includes large-Q surface phonon \cite{aref2016quantum,PhysRevX.5.031031} and bulk phonon \cite{chu2018creation} cavities, and emitters of non-classical acoustic waves \cite{Chu199}, all of which are being optimized to increase the precision of control by engineering optomechanical \cite{maccabe2019phononic} and piezoelectric interactions \cite{PhysRevX.8.031007}.

A natural next step towards expanding this toolbox is to engineer a more efficient coupling between optically controllable emitters of single phonons --- for example negatively charged nitrogen-vacancy defects in diamond (NV$^-$) --- and acoustic resonators \cite{jayich2017jopt}. This strain coupling is predominantly determined by a cavity's ability to spatially confine phonons beyond the diffraction-limited mode volume $\propto \lambda^3$, and to suppress the dissipation of phonons into radiative and non-radiative channels. These effects, quantified by the effective mode volume of the acoustic mode \Veff, and acoustic quality factor $Q$ respectively, provide a measure of cavity performance through the acoustic analogue of the Purcell factor $P_F\propto Q/\Veff$ \cite{schmidt18prl}.

Analogous challenges in electrodynamics are typically addressed by optimizing either one of the characteristics determining the Purcell factor. In plasmonic systems, large $P_F\sim 10^5$ \cite{kongsuwan2017suppressed,chikkaraddy2016single} can be realized in systems with very small effective mode volumes $\Veff\sim 10^{-4} \lambda^3$, and small $Q\sim 10$ \cite{koenderink2017single}. Alternatively, all-dielectric systems \cite{lodahl2015interfacing} can provide large $Q$ through the engineering of near perfect reflection by Bragg mirrors \cite{reithmaier2004strong} or by utilizing whispering gallery modes \cite{aoki2006observation}, but they exhibit no ability to spatially confine light below the diffraction limit. Recently reported approaches are merging these advantages, either by utilizing hybrid plasmonic-dielectric cavities \cite{doeleman2016antenna,bozzola2017hybrid}, or implementing defects in photonic crystal waveguides (with intrinsically large $Q$) specifically engineered to strongly localize the electric field (with $\Veff\sim10^{-5}\lambda^3$) \cite{weiss16acsphot,englund17prl,weiss18sciadv}.

\begin{figure}[htbp!]
    \centering
    \includegraphics[width=.8\linewidth]{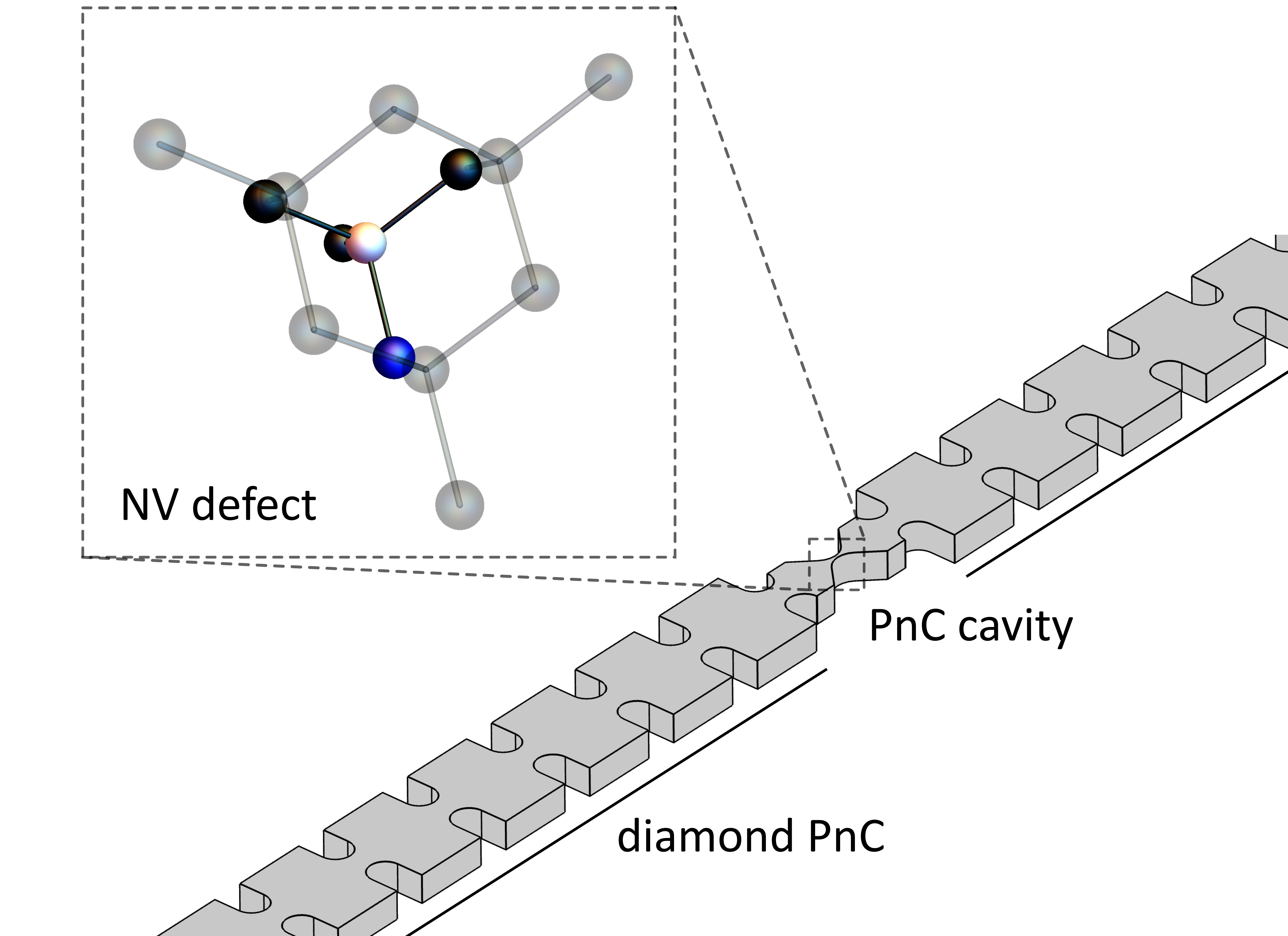}
    \caption{Schematic of an acoustic diamond resonator with ultra-small volume mode localized at the defect (see Section~\ref{sec:local_cavity}) of a finite 1D phononic waveguide (see Section~\ref{sec:bragg_mirrors}). Orbital states of the nitrogen-vacancy defect (NV$^-$) are coupled to the strain of the mode via the parametric and resonant interactions (see Section~\ref{sec:coupling}).}
    \label{fig:Fig1}
\end{figure}

In this work, we translate this last concept into the acoustic domain, by designing sub-wavelength defects in diamond phononic crystal waveguides (PnCW), schematically shown in Fig.~\ref{fig:Fig1}. The central defect localizes the strain through an acoustic analogue of the lightning-rod effect \cite{van1991singular}, confining the energy of the acoustic mode into effective volumes a few orders of magnitude below the diffraction-limited $\lambda^3$. As we show below, a significant localization characterized with $\Veff\sim 10^{-4}\lambda^3$ can be achieved in diamond waveguides. This leads to a significant enhancement of the coupling between phonon emitters in diamond (e.g. orbital states of NVs), and mechanical modes of the structure, opening pathways to implementing high-cooperativity NV-phonon coupling on the nanoscale. In particular, we estimate that in our systems the cooperativity of both the resonant and parametric couplings, which we describe in more detail in Section~\ref{sec:coupling}, can be enhanced to $C\sim 8$ and $C\sim 0.7$, respectively.

The paper is structured as follows: in Sections~\ref{sec:local_cavity} and \ref{sec:bragg_mirrors} we introduce two key elements of our design --- mechanisms of strain localization in ultra-small mode volumes, and designs of crystal waveguides with complete acoustic bandgaps. In Section~\ref{sec:entire_system} we assemble these two elements into Phononic Crystal Waveguide (PnC) cavities, and provide estimates for their effective mode volumes. Finally, in Section~\ref{sec:coupling} we discuss two mechanisms of coupling between the intrinsic strain fields of the acoustic modes and the orbital states of the NV centers, calculate the coupling strengths, resulting cooperativities, and efficiencies of the resonant and off-resonant cooling protocols \cite{rabl13prl,jayich16prapplied,fuchs18prl,jayich2019qscitech}.

\section{Cavity design}
\label{sec:local_cavity}

Let us consider a modal picture of the elastic response of an arbitrary mechanical resonator. The confinement of a particular elastic mode (indexed as $\alpha$) can be qualitatively expressed through an effective volume \Veffa~which, neglecting the spectral dispersion and losses in the acoustic response of the materials, can be defined as \cite{eichenfield2009nature}
\begin{equation}\label{Veff.def}
    \Veffa = \frac{\int \text{d}\mathbf{r} ~h(\mathbf{r};\alpha)}{\text{max}_\mathbf{r}~h(\mathbf{r};\alpha)},
\end{equation}
where the local energy density $h(\mathbf{r};\alpha)$, averaged over the acoustic period $2\pi/\Omega_\alpha$, is given as a sum of the strain and kinetic energy densities: 
\begin{equation}\label{h.def}
    h(\mathbf{r};\alpha) = \frac{1}{2}\etensor_\alpha(\mathbf{r})^*:\ctensor(\mathbf{r}):\etensor_\alpha(\mathbf{r}) + \frac{1}{2}\omega_\alpha^2\rho(\mathbf{r})|\mathbf{u}_\alpha(\mathbf{r})|^2.
\end{equation}
The mode is characterized by the displacement field $\mathbf{u}_\alpha$ and strain tensor defined as its symmetrized gradient $\etensor_\alpha=\nabla_S \mathbf{u}_\alpha$ \cite{auld1973acoustic}. The local density $\rho$ and the 4-th rank stiffness tensor $\ctensor$ are treated as parameters. When integrated over the volume of the resonator, the two terms in Eq.~\eqref{h.def} yield equal contributions to the total energy. However, we should note that this equivalence does not hold locally. 

\begin{figure}[htbp!]
    \centering
    \includegraphics[width=\linewidth]{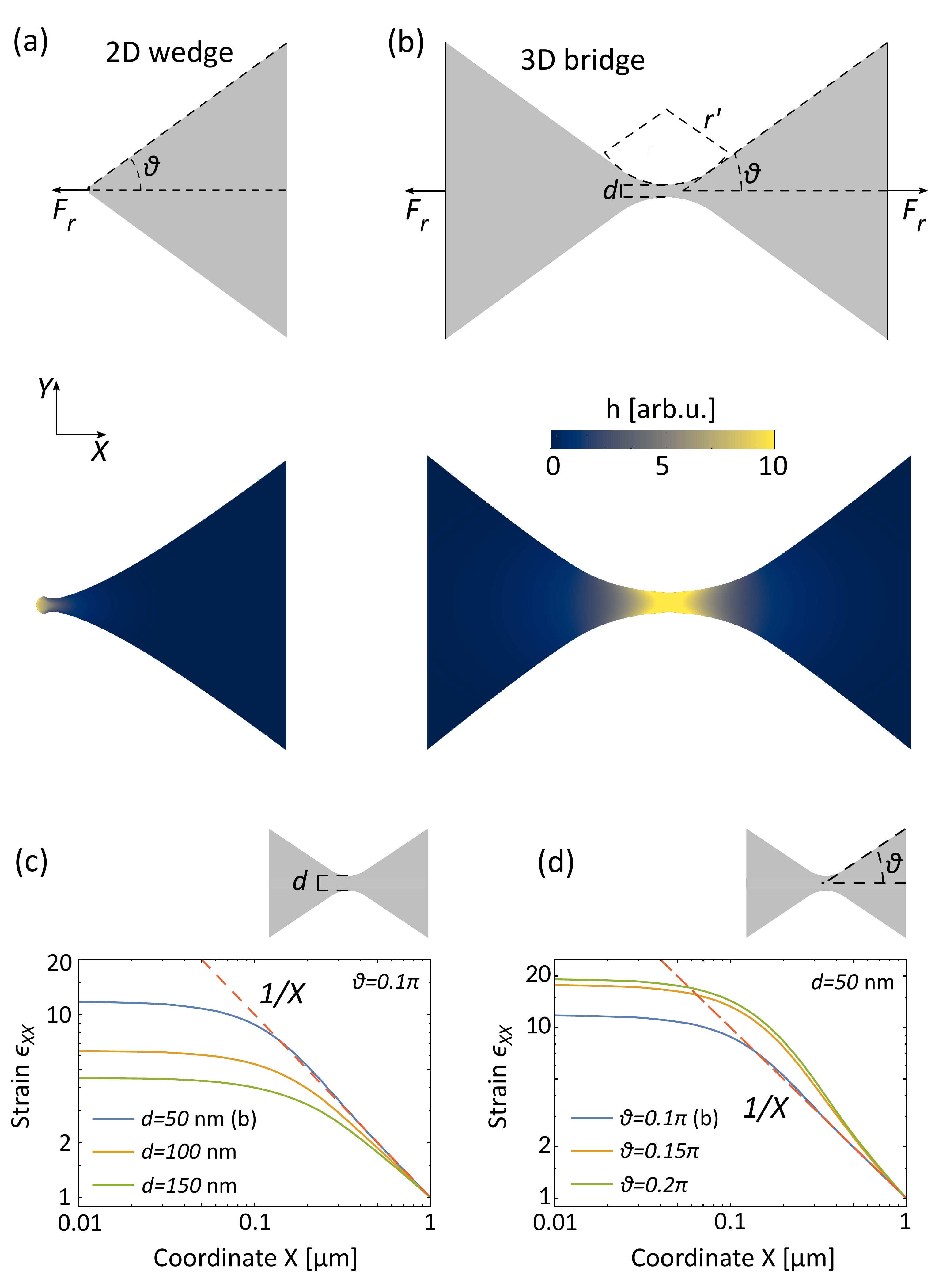}
    \caption{Mechanism of non-resonant strain localization in a (a) truncated 2D wedge and (b) 3D bridge. The two structures, with geometries depicted with gray schematics in the top panels (note that for the 3D structure we show the cross section structure schematic and energy distribution in the $\hat{X}\hat{Y}$ symmetry plane), are deformed in the lower panels according to the displacement field, and colored to indicate the energy density distribution $h(\mathbf{r}; \alpha)$. (c,d) Dependence of the axial strain on parameters of the 3D bridge marked schematically in (b): bridge thickness $d$ and opening angle $\theta$ calculated along the symmetry axis $X$ (dashed horizontal line at $Y=0$). The red dashed line represents the $X^{-1}$ strain decay found for a 2D wedge, indicative of the acoustic analogue of the lightning rod effect. The bridge is engineered in a slab of thickness $t=0.5~$\textmu m used throughout this work, and the curvature is set to $r'=0.6~$\textmu m. Here we denote laboratory frame coordinates as $XYZ$ to differentiate from the $xyz$ coordinate system related to the orientation of NV$^-$ color centres.}
    \label{fig:FigMechanism}
\end{figure}

From the definition of the effective mode volume, we find that \Veffa~can be reduced by locally enhancing the energy density $h$. Notably, this local enhancement does not have to originate from any resonant phenomena, and --- within the approximation of dispersionless material properties, which is largely correct in GHz acoustics --- can be inferred from static analysis. This is particularly interesting given the pivotal role non-resonant mechanisms have played in the development of state-of-the-art optical resonators, such as plasmonic picocavities~\cite{benz} and dielectric subwavelength resonators \cite{weiss16acsphot,englund17prl,weiss18sciadv}. Here we investigate a new type of non-resonant acoustic localization effect --- the acoustic analogue of the lightning-rod effect, that localizes the strain in a tapered, sub-wavelength bridge structure shown schematically in Fig.~\ref{fig:FigMechanism}(b). To illustrate the fundamental characteristics of this mechanism, we can consider a simplified 2D system, with the tapered semi-infinite diamond wedge shown in Fig.~\ref{fig:FigMechanism}(a). When an axial force $\mathbf{F_r}=F_r \hat{X}$ (with $F_r<0$) is applied to its tip, the strain becomes localized and exhibits a $r^{-1}$ divergence \cite{landaubook}:
    \begin{equation}
            \varepsilon_{rr}(r,\phi) = -\frac{F_r}{E}\frac{\cos\phi}{r\left[\theta-\frac{1}{2}\sin(2\theta) \right]},        
    \end{equation}
    \begin{equation}
        \varepsilon_{\phi r}(r,\phi)=\varepsilon_{\phi \phi}(r,\phi)=0,
    \end{equation}
where $r$ ($r=0$ at the tip) and $\phi$ are the polar coordinates, and $E$ is the Young's modulus of the material. Here we approximate the diamond as an isotropic medium with $E=1050$~GPa, Poisson ratio $\nu=0.2$ and density $\rho=3500$~kg/m$^3$ \cite{PhysRevLett.120.213603}. The energy density and displacement shown in Fig.~\ref{fig:FigMechanism}(a) were calculated using COMSOL Multiphysics\textsuperscript{\textregistered} software~\cite{comsol} assuming a finite tip width of 10~nm, to ensure that the problem is well-posed, and therefore exhibits finite localization near the tip.

Building on this phenomenon of a lightning-rod-like behavior, we consider the tapered bridge as a symmetric, 3D finite-width extension of the tip setup with thickness (along the out-of-plane axis $Z$) $t=0.5~$\textmu m, and anticipate a similar localization of the axial strain at the center of the bridge (see Fig.~\ref{fig:FigMechanism}(b)). To illustrate this effect quantitatively, in Fig.~\ref{fig:FigMechanism}(c,d) we plot the axial ($XX$) component of the strain field along the axis of the bridge (where polar coordinate $r$ becomes $X$) for a range of (c) bridge widths $d$ and (d) taper angles $\theta$. All the values in plots are normalized to the strain at $X=1~$\textmu m for clarity. We find that the largest localization of strain is offered by the narrowest bridges, and larger wedge angles $\theta$ --- this latter behavior being a deviation from the one-sided analytical system. In agreement with the analytical model of a one-sided wedge, the strain exhibits a $X^{-1}$ decay away from the bridge, and becomes effectively homogeneous inside the bridge structure. We can therefore estimate that the mode is localized in the 2D plane to an approximate area of the bridge $r' d$, or in 3D to its volume $t r'd$. 

Local strain could be further enhanced (and effective surface and volume --- reduced) if we considered an even narrower bridge or a smaller curvature radius $r'$. However, we impose lower bound on $d$ corresponding to the state of the art of lithography techniques in diamond \cite{dory2019inverse}, where the smallest reliably fabricated features are about 50~nm. Simultaneously, we keep the curvature $r'$ --- which determines the length of the bridge --- larger, to accommodate multiple NVs which would be homogeneously coupled to the cavity mode. We briefly discuss collective coupling effects in Appendix~\ref{Appendix.CollectiveEffects}.

\section{Quasi-1D phononic crystal}
\label{sec:bragg_mirrors}

\begin{figure}[htbp!]
\centering
\includegraphics[width=\linewidth]{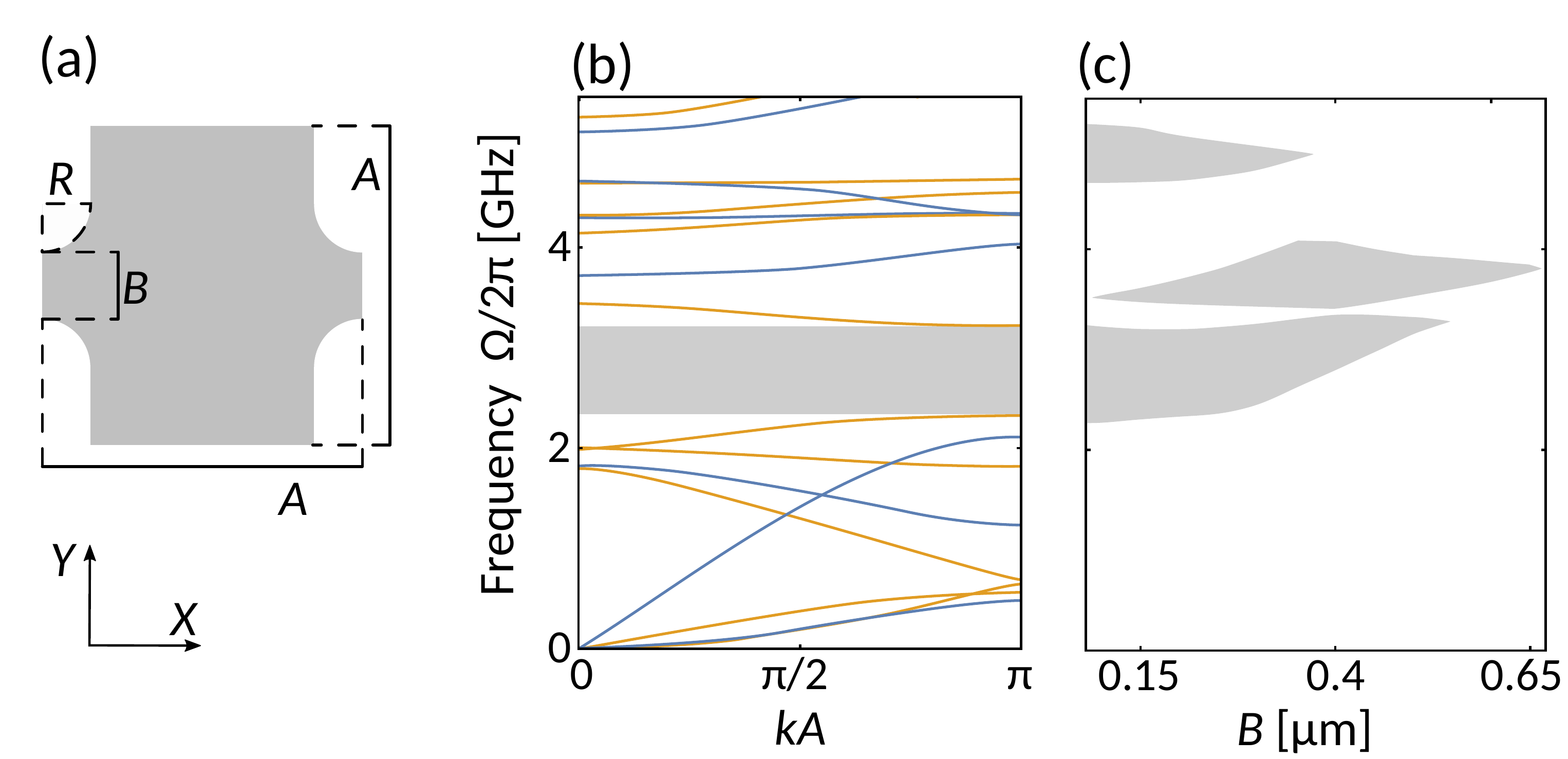}
\caption{Engineering a quasi-1D phonon crystal in a $t=0.5~\mu$m thick slab of diamond. (a) Geometry of a unit cell with parameters given in the text. (b) Phononic dispersion with $Z$-symmetric modes, with $u_Z$ component of the symmetric with respect to the $Z$ plane symmetry, is shown in blue lines. Dispersion of the similarly defined $Z$-antisymmetric modes is depicted with orange lines. Modes of both symmetries exhibit a broad complete acoustic bandgap (shaded region), (c) which can be tuned or closed by changing the width of the connecting structures $B$. Geometric parameters are given in the text. }
\label{fig:FigPhononShield}
\end{figure}

While the acoustic lightning-rod effect provides a mechanism for non-resonant sub-wavelength localization of the strain field, the modal properties of the cavity mode $\alpha$ in a phononic crystal waveguide --- its resonant frequency $\omega_\alpha$ and quality factor $Q_\alpha$ --- are determined by the reflection from the acoustic crystal waveguide structure (see schematics in Fig.~\ref{fig:Fig1}), and the larger-scale structure of the cavity.

In Fig.~\ref{fig:FigPhononShield}(a) we present a design for a unit cell of a 1D phononic crystal waveguide based on the designs previously used for 2D phononic shields in silicon \cite{painter2019arxiv,chan12apl}, and --- more recently --- in diamond \cite{jayich2019qscitech}. The geometric parameters shown in Fig.~\ref{fig:FigPhononShield}(a), in particular the width of the connecting bridge $B$, govern the bandgap of the structure along the $X$ direction. By choosing $(A,B,R)=(1.925,0.2,0.29)~\mu$m, we engineer the dispersion relation of the $Z$-symmetric and $Z$-antisymmetric modes (see the used definition of symmetry, and blue and orange lines in Fig.~\ref{fig:FigPhononShield}(b)) to exhibit a complete bandgap between $2.3$ and $3.2$~GHz. All the results were obtained with COMSOL~\cite{comsol} by implementing Floquet boundary conditions along the axis of the unit cell.

We should also note that, unlike in photonic crystal waveguide cavities, acoustic cavities can be engineered as a single defect in a phononic crystal waveguide with a complete bandgap, without any adiabatic transition region between the defect and the Bragg mirrors. This is because phonons emitted from the cavity cannot efficiently outcouple into free radiation in the surrounding medium, and the changes to the phonon momentum at the interface between the cavity and the acoustic Bragg mirror can be  arbitrarily large. Therefore, using the designs of unit cell given in Fig.~\ref{fig:FigPhononShield}(a), we can proceed to interface the Bragg structure directly with the defect cavity, and calculate the effective volumes and quality factors of the resulting cavity modes.

\section{Phononic crystal waveguide cavity}
\label{sec:entire_system}

\begin{figure}[htbp!]
    \centering
    \includegraphics[width=\linewidth]{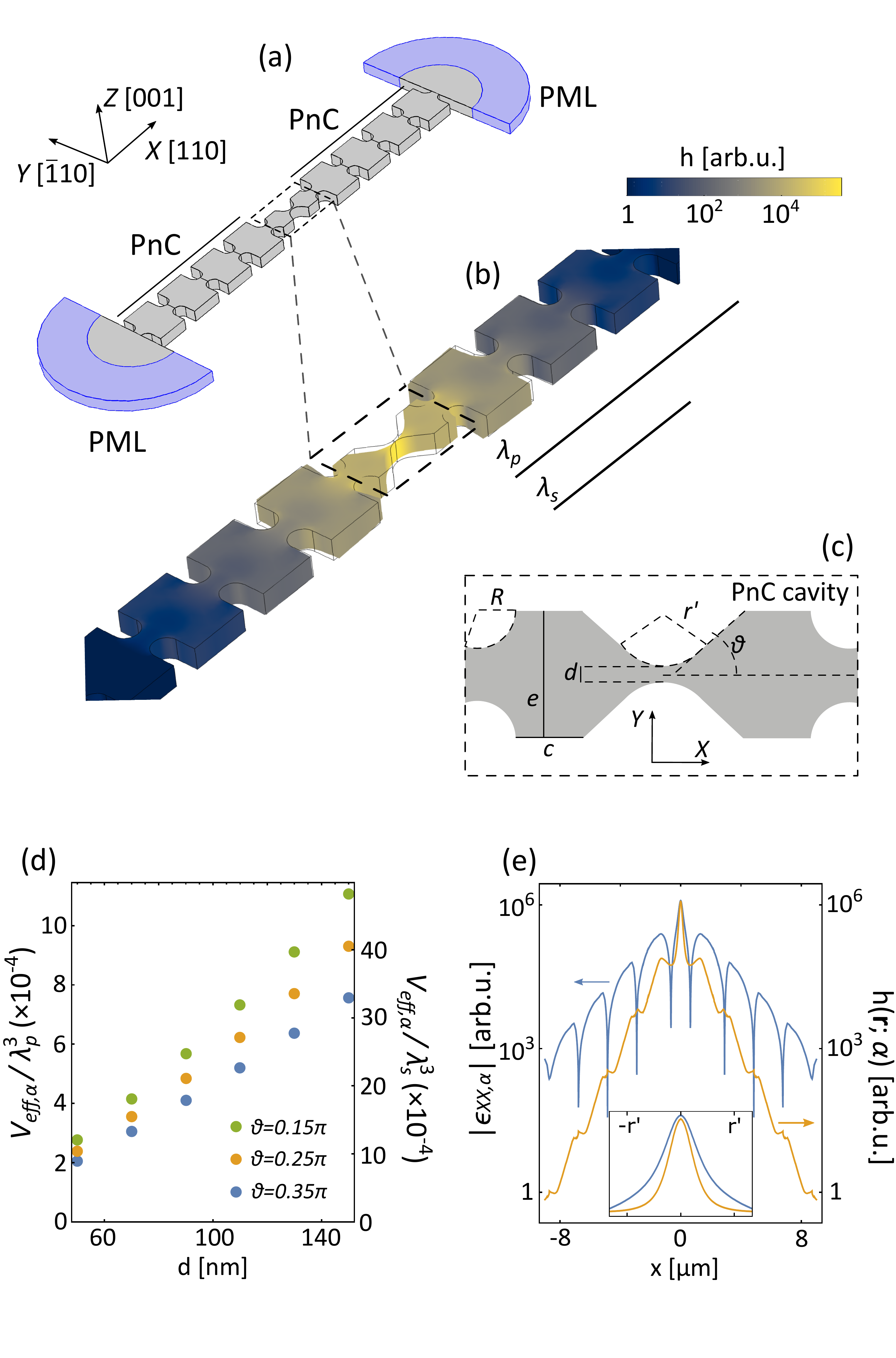}
    \caption{Design and characteristics of the PnCW cavity. (a) Complete design of the cavity and anchoring to the substrate (colored regions denoting Perfectly Matching Layers (PML)), with the (b) magnified region showing the geometry of the interface between the defect and a unit cell of the 1D phononic crystal. As in the reported experimental realizations of diamond waveguides \cite{jayich16prapplied}, we assume that the crystal coordinate system $(X,Y,Z)$ is chosen so that $X\parallel[110]$, $Y\parallel[\bar{1}10]$, $Z\parallel[001]$. Bars underneath denote longitudinal $p$ and shear $s$ elastic wavelengths in diamond at the mode frequency $\Omega_\alpha/2\pi=2.838$~GHz. In (c), as in Fig.~\ref{fig:FigMechanism}, we show the displacement field of the cavity mode and energy density distribution $h(\mathbf{r}; \alpha)$, calculated for $(d,c,e,r',R)=(50,400,960,375,290)~$nm and $\theta=0.15\pi$. (d) Dependence of the effective mode volume \Veffa, normalized by the wavelength of the $p$ wave in diamond, on the thickness of the cavity bridge $d$ and opening angle $\theta$. (e) Energy density and strain $|\varepsilon_{XX,\alpha}|$ amplitude along the axis of the crystal demonstrates the exponential localization of the mode to the cavity. Inset shows the distribution of $|\varepsilon_{XX,\alpha}|$ and $h(\mathbf{r}; \alpha)$ in the central part of the cavity with linear vertical scale.}
    \label{fig:FigFinalCavity}
\end{figure}

In Fig.~\ref{fig:FigFinalCavity}(a) we present the design and characteristics of the entire PnC cavity. The dimensions of the bridge structure in the cavity (its width $d$, curvature radius $r'$ and opening angle $\theta$; see Fig.~~\ref{fig:FigFinalCavity}(c)) determine the localization of elastic strain energy, and geometric parameters $b$ and $c$ govern the frequency of the cavity mode. When this frequency matches  the bandgap of the quasi-1D acoustic crystal, radiative phonon dissipation is suppressed. In particular, in panel (e) we show the confinement of the mode tuned to the centre of the bandgap (parameters are given in the caption) with $\Omega_\alpha/2\pi=2.838~$GHz mode. For the finite phononic crystal spanning 4 unit cells on each side of the cavity, we calculate the acoustic quality factor to reach $Q\sim 10^6$. For this structure, we find effective mode volumes \Veffa~of the order of $10^{-4}\lambda_p^3$, or $10^{-3}\lambda_s^3$, where $\lambda_p=\sqrt{E(1-\nu)/[\rho(1+\nu)(1-2\nu)])}$ and $\lambda_s=\sqrt{E/[2\rho(1+\nu)])}$ are the longitudinal and shear wavelengths of elastic wave in bulk diamond at the frequency of the mode.

In these calculations, the only mechanisms limiting the mechanical quality factor $Q$ are related to the radiative dissipation and clamping losses, while contributions from intrinsic mechanisms are neglected. We briefly discuss the state of the art of crystal cavity fabrication in diamond, and methods of mitigating these limitations, in Section~\ref{section:fab}. Furthermore, throughout the rest of the paper, we take a more conservative estimate of the mechanical quality factor $Q=10^5$.

\section{Orbital states of NV and coupling to the strain}
\label{sec:coupling}

\begin{figure}[htbp!]
\centering
\includegraphics[width=\linewidth]{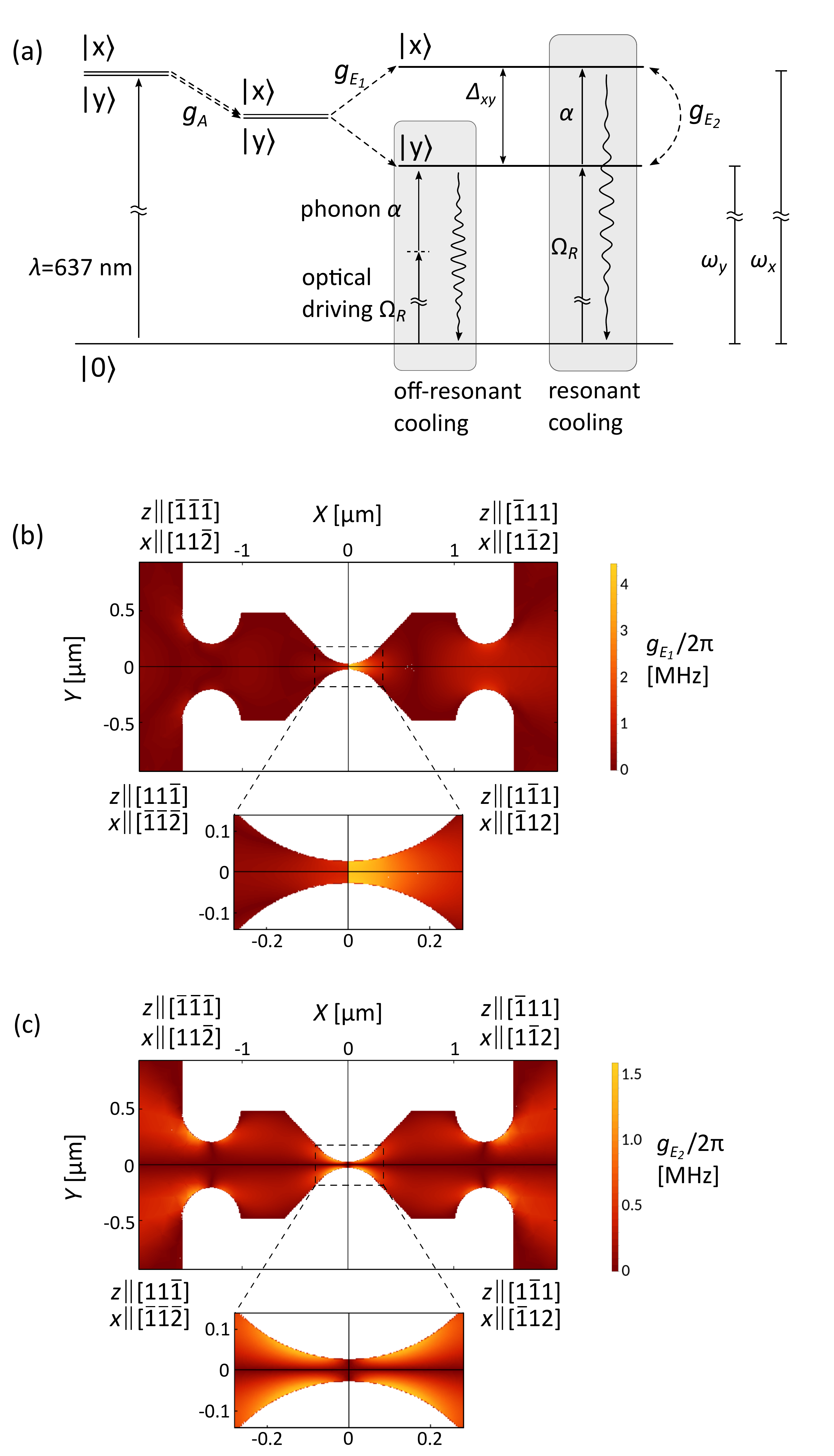}
 \caption{Coupling between orbital states of the NV and strain field. 
 (a) Energy structure of the $m_s=0$ manifold of NV defects in diamond and effects of coupling with the crystal strain, as discussed in the text. Two cooling protocols proposed by Kepesidis \textit{et al.} \cite{rabl13prl}, involving optical driving of the excited states with Rabi frequency $\Omega_R$, and engineering coupling with GHz phonons in mode $\alpha$, are schematically shown in shaded rectangles. (b,c) Position-dependent coupling parameters (b) $g_{E_1}$ and (c) $g_{E_2}$ calculated for NV centers 5~nm below the surface of the diamond, with different orientations of the NV coordinate system (orientation of the $z$ axis is given in the corners by Miller indices).}
\label{fig:FigCouplings}
\end{figure}

We now consider coupling between a phonon emitter --- in this case, a negatively charged NV$^-$ center positioned at the center of the cavity --- and the strain field of the acoustic cavity mode. At the microscopic level, the modal strain induces displacement of the atoms making up the NV, which in turn modifies the Coulomb interaction between the ions and electrons of the NV \cite{doherty2013nitrogen}, shifting and mixing its energy levels. This interaction is to a good approximation linear in strain, and can be thus written in a general form as
\begin{equation}
    H_\text{strain-NV} = \sum_{i,j \in \{x,y,z\}} V_{ij} \epsilon_{ij},
\end{equation}
where the $V_{ij}$ operators describe transitions between states of the NV$^-$, and $\epsilon_{ij}$ is the strain tensor in index notation. This general expression can be rewritten in a more convenient basis which reflects the $C_{3v}$ symmetry of the orbital wavefunctions of the NV (see e.g. PhD thesis by Lee \cite{lee} for an excellent introduction to the subject and the formalism). The projection onto the irreducible representations $\Gamma \in \{A,E_1,E_2\}$ of the $C_{3v}$ group allows us to separate the contributions from three interactions \cite{jayich2017jopt}:
\begin{equation}\label{Ham.strains}
    H_\text{strain-NV} = H_{A} +H_{E_1} +H_{E_2}.
\end{equation}
Here we consider the interaction between strain and orbital states within the excited $^3E$, $m_s=0$ manifold of the NV$^-$, which exhibit much stronger strain-orbit coupling than those in the ground state $A_2$ (identified as the orbital-singlet state with $A_2$ symmetry and triplet spin component) \cite{jayich2017jopt}. Since we consider only $m_s=0$ manifold, in the following discussion we simplify the notation by omitting the spin degree of freedom. In the absence of strain, $^3E$ is a doublet of degenerate molecular orbital states denoted as $\ket{x}$ and $\ket{y}$, which describe electron configurations of 6 electrons (or equivalently 2 holes) of the NV$^-$ occupying single electron orbitals $a_1(1)$, $a_1(2)$, $e_x$ and $e_y$. In both $\ket{x}$ and $\ket{y}$, 4 electrons occupy the lowest-energy $a_1(1)$ and $a_1(2)$ orbitals. For $\ket{x}$, two remaining electrons are both either in $e_x$ (resulting in the two holes occupying $a_1(2)$ and $e_y$ orbitals: $\ket{x} = \ket{a_1(2)e_x-e_xa_1(2)}$). Conversely, in $\ket{y}$, the electrons occupy $e_y$ ($\ket{y} = \ket{a_1(2)e_y-e_ya_1(2)}$)\cite{lee}. These orbital states are defined in the local coordinate system of the NV, where axis $z$ is chosen along the N-V direction (aligned with one of the $[\bar{1}\bar{1}\bar{1}]$, $[11\bar{1}]$, $[\bar{1}11]$, or $[1\bar{1}1]$ crystallographic directions), and are thus fixed unambiguously for a specific NV. The axis $x$ is determined by the projection of any one of the three vacancy-carbon directions onto the plane perpendicular to $z$. This freedom of choice of the local coordinate system has led to some confusion in the literature, which we aim to clarify below and, in more detail, in Appendix~\ref{appendix.orientations}. 

It can be shown that the three terms in the Hamiltonian given in Eq.~\eqref{Ham.strains} couple states $\{\ket{x},\ket{y}\}$ with the elements of the strain tensor $\etensor$ as
\begin{equation}
    H_A = \big[\lambda_{A}{\varepsilon}_{zz}+\lambda_{A'}({\varepsilon}_{xx}+{\varepsilon}_{yy})\big]\big(\ket{x}\bra{x}+ \ket{y}\bra{y}\big),
\end{equation}
\begin{equation}\label{H1}
    H_{E_1}=\big[\lambda_E({\varepsilon}_{yy}-{\varepsilon}_{xx})+2\lambda_{E'}{\varepsilon}_{xz}\big] \big(\ket{x}\bra{x}-\ket{y}\bra{y}\big),
\end{equation}
\begin{equation}\label{H2}
    H_{E_2} = 2\big[\lambda_E {\varepsilon}_{xy}+\lambda_{E'}{\varepsilon}_{yz}\big]\big(\ket{x}\bra{y}+  \ket{y}\bra{x}\big),
\end{equation}
where all the strain tensor elements are taken at $\mathbf{r}_{\text{NV}}$ the position of the NV centre, and in the local coordinate system $xyz$ determined by the orientation of the NV. Transformation from the laboratory frame of reference $XYZ$ to that of the NV is briefly described in Appendix~\ref{Appendix.AxisRotation}. $\lambda_A$, $\lambda_{A'}$, $\lambda_E$, and $\lambda_{E'}$ are strain susceptibilities or orbital-strain coupling constants of the NV \cite{jayich16prapplied}.
Effects of the \textit{static} strain (either external or intrinsic), originating from each of these terms, is schematically shown in Fig.~\ref{fig:FigCouplings}(a). As strain $E_{1}$ lifts the degeneracy of $\{\ket{x},\ket{y}\}$ states, we 
can introduce physically meaningful states $\{\ket{x}_s,\ket{y}_s\}$ defined unambiguously as eigenstates of the Hamiltonian $H_{\text{strain-NV}}$ under static strain. 

In this work we do not consider these static mechanisms in any more detail, but simply assume that the static strain lifts the degeneracy of the orbital states, and allows us to define a physically-relevant coordinate system $x_sy_sz$ corresponding to the $\{\ket{x}_s,\ket{y}_s\}$ states.

\subsection{Coupling to the acoustic cavity mode}
\label{subsec:acoustic.strain}

We can now consider the coupling between the orbital states of the NV$^-$ and dynamic, acoustic and quantized modes of the cavity.
To this end, we introduce a quantized picture of the elastic vibrations and then consider the exact form of coupling between the NVs and the mode of the resonator. The quantization of the elastic field can be carried out following the scheme previously explored for optical subwavelength lossless resonators with inhomogeneous field distribution \cite{esteban2014strong}. We outline this procedure in Appendix~\ref{Appendx.Quantization}, and arrive at the quantum operators corresponding to the displacement field $\hat{\mathbf{u}}_\alpha$ and strain tensor $\hat{\etensor}_\alpha$, the latter of which takes the following form:
\begin{equation}\label{etensor.def}
    \hat{\etensor}_\alpha(\mathbf{r}) = \sqrt{\frac{\hbar }{2\Omega_\alpha}}\left\{\etensor_\alpha(\mathbf{r}) b_\alpha + \left[\etensor_\alpha(\mathbf{r})\right]^* b_\alpha^\dag\right\}.
\end{equation}
In the above definition, the classical strain tensor $\etensor_\alpha$ is normalized as $\Omega_\alpha^{-2}\int \etensor_\alpha(\mathbf{r}):\ctensor(\mathbf{r}):\etensor_\alpha(\mathbf{r})~\text{d}\mathbf{r} = 1$, and phonon annihilation and creation operators $b_\alpha$ and $b_\alpha^\dag$ follow the bosonic commutation relations $\left[b_\alpha,b_{\alpha'}^\dag\right]=\delta_{\alpha,\alpha'}$.

We should note that the above quantization procedure is not exact for any realistic system with non-vanishing radiative dissipation of the acoustic waves. Similar to the case of photonic crystals or scattering particles, the integral given in Eq.~\eqref{Veff.def} diverges as the integration volume is increased due to the radiative component of the fields, and we should embrace the picture of acoustic analogues of the quasi-normal modes \cite{Kristensen2014,Sauvan13}. However, for the high-Q modes discussed here the corrections to the coupling parameters or the quality factors should be negligible.


\subsection{Parametric coupling ($E_1$)}
\label{subsec:parametric}

We first consider the parametric coupling between the excited orbital states of the NV and the strain field. This term results from the interaction between strain and molecular orbitals which both transform as $E_1$ irreducible representation of the $C_{3v}$ symmetry group of the NV, and is described by the quantized version of the Hamiltonian given in Eq.~\eqref{H1}:
\begin{equation}
    H_{E_1} = [\lambda_E(\hat{\varepsilon}_{yy}-\hat{\varepsilon}_{xx})+\lambda_{E'}(\hat{\varepsilon}_{xz}+\hat{\varepsilon}_{zx})]\left(\ket{x}\bra{x}-\ket{y}\bra{y}\right),
\end{equation}
Using the definition in Eq.~\eqref{etensor.def}, 
we can rewrite it in terms of the effective strain-orbit coupling with coupling coefficient $g_{E_1}$:
\begin{equation}
    H_{E_1} = g_{E_1}(b_\alpha+b_\alpha^\dag)\left(\ket{x}\bra{x}-\ket{y}\bra{y}\right),
\end{equation}
where $g_{E_1}=\sqrt{\hbar /2\Omega_\alpha}\{\lambda_E[\varepsilon_{\alpha,yy}(\mathbf{r}_{\text{NV}})-\varepsilon_{\alpha,xx}(\mathbf{r}_{\text{NV}})]+2\lambda_{E'}\varepsilon_{\alpha,xz}(\mathbf{r}_{\text{NV}})\}$. Thus $g_{E_1}$ implicitly depends on the orientation of the NV through the expression of the strain tensor in the NV coordinate system. Since $\lambda_E\gg \lambda_{E'}$ (we take $\lambda_E= -0.85~$PHz and $\lambda_{E'}= 0.02~$PHz \cite{jayich16prapplied}), the coupling term will be dominated by the diagonal elements of the strain tensor $\varepsilon_{\alpha,xx}$ and $\varepsilon_{\alpha,yy}$, and therefore by the longitudinal components of the strain in these coordinates. 

In the map of $g_{E_1}$ in Fig.~\ref{fig:FigCouplings}(b) we consider separately four orientations of the NV, defined by the $z$ ($x$) axes along the $[\bar{1}\bar{1}\bar{1}]$ ($[{1}{1}\bar{2}]$), $[11\bar{1}]$ ($[\bar{1}\bar{1}\bar{2}]$), $[\bar{1}11]$ ($[{1}\bar{1}{2}]$), or $[1\bar{1}1]$  ($[\bar{1}{1}{2}]$) directions \footnote{Note that the orientations of the $x$ are chosen by arbitrarily selecting a carbon atom which defines it (see discussion in Appendix~\ref{appendix.orientations}).}, in a plane 5~nm below the upper surface of the diamond. Such shallow defects can be generated using low-energy ion implantation \cite{smith2019colour}. However, we should note that unlike surface acoustic waves, or the strongly localized modes of a triangular cross-section PnCW discussed by Meesala \textit{et al.} \cite{PhysRevB.97.205444} (dubbed \textit{flapping modes}; see discussion in Section~\ref{SiV}), our cavity exhibits an approximately constant strain along its depth (Z axis), and the exact positioning of the NV can be optimised to shield the defect from external electric fields or surface strain. 

The calculated value of coupling $g_{E_1}$ becomes more meaningful if we compare it to the dephasing rates of the involved electronic states $\Gamma_{x/y}$ ($\Gamma_{x/y}/2\pi \approx 15~$MHz~\cite{hanson10prl}), and the re-thermalization rate of cavity phonons $\gamma_{\text{th}}=n_\text{th}\Omega_\alpha/(2Q)$ (with $n_\text{th}=[\exp(\hbar \Omega_\alpha/k_B T)-1]^{-1}$). For the calculations carried out in this section, we consider resonators with slightly lower, more realistic mechanical quality $Q=10^5$, operating at temperature of 4~K (where $n_\text{th}(\Omega_\alpha/2\pi=2.4~\text{GHz})\approx 35$). We can then calculate the parametric coupling cooperativity $C_{E_1}=4g_{E_1}^2/(\gamma_{\text{th}} \Gamma_{x/y})$ for the maximum coupling $g_{E_1}/2\pi=5$~MHz as reaching $C_{E_1}\approx 8$, suggesting that the system can reach a high-cooperativity regime.

Parametric coupling also offers a pathway to implementing an off-resonant phonon cooling protocol, as proposed by Wilson-Rae \textit{et al.} in Ref.~[\citenum{wilsonrae04prl}] (see Fig.~\ref{fig:FigCouplings}(a) for a schematic of the protocol). In this protocol, the NV is excited from its ground electronic state $\ket{0}$, optically driven with Rabi frequency $\Omega_R$ (here we put $\Omega_R/2\pi =15~\text{MHz}=\Gamma_{x/y}/2\pi$ to saturate the electronic states \cite{rabl13prl}) to a virtual state $\omega_y-\Omega_\alpha$ which is red-detuned from  the lower energy orbital state $\ket{y}$ by the frequency $\Omega_\alpha$ of the acoustic cavity. The NV subsequently absorbs the cavity phonon $\alpha$, and relaxes to the ground state $\ket{0}$ by optical emission at $\omega_y$. This cycle cools the acoustic mode $\alpha$ at a rate \cite{rabl13prl} $\Gamma_{E_1} = g_{E_1}^2\Omega_R^2/(\Gamma_{x/y} \Omega_\alpha^2)\approx2\pi \times 60~\text{Hz}$. As the cooling rate is much lower than the re-thermalization rate $\gamma_{\text{th}}$, the off-resonant scheme cannot efficiently cool the mechanical vibrations of the resonator. A similar conclusion was found for a submicron, high-Q acoustic cantilever resonator hypothesized by Kepesidis \textit{et al.} \cite{rabl13prl}.

Finally, we should point our that in the above formulation, the values of coupling coefficient $g_{E_1}$ explicitly depend on the choice of the local $x$ axis (see coupling maps in Fig.~\ref{fig:FigCouplingsOtherC}(a)). To remedy this non-physical effect, we need to account for the presence of the static strain which reduces the symmetry of the NV system, and express the dynamical coupling (both parametric and resonant, discussed in the following subsection), in the basis of eigenstates of the static-strained NV $\{\ket{x}_s,\ket{y}\}$. WE include a more detailed formulation of this method in Appendix~\ref{appendix.orientations}

\subsection{Resonant coupling ($E_2$)}
\label{subsec:resonant}

In the Hamiltonian given in Eq.~\eqref{Ham.strains}, the only term describing resonant transitions between excited orbital states $\ket{x}$ and $\ket{y}$ is associated with the components of the strain tensor and molecular orbitals which both transform as $E_2$ irreducible representations of the $C_{3v}$ group:
\begin{equation}
    H_{E_2} = [\lambda_E(\hat{\varepsilon}_{xy}+\hat{\varepsilon}_{yx})+\lambda_{E'}(\hat{\varepsilon}_{yz}+\hat{\varepsilon}_{zy})]\left(\ket{x}\bra{y}+\ket{y}\bra{x}\right).
\end{equation}
As above, we can write down the quantized version of this interaction
\begin{equation}
    H_{E_2} = g_{E_2}\left(\ket{x}\bra{y}+\ket{y}\bra{x}\right)(b_\alpha^\dag + b_\alpha),
\end{equation}
finding $g_{E_2}=2\sqrt{\hbar /2\Omega_\alpha}[\lambda_E\varepsilon_{\alpha,xy}+\lambda_{E'}\varepsilon_{\alpha,yz}]$.

Values of the $g_{E_2}$ coupling are shown in Fig.~\ref{fig:FigCouplings}(c) for the four NV orientations. Unlike in the case of the parametric interaction, the maximum coupling is not found for the emitter placed in the narrowest part of the bridge, but rather near its edges. This is because $g_{E_2}$ depends predominantly on the off-diagonal components of the strain tensor $\varepsilon_{\alpha,xy}=\varepsilon_{\alpha,yx}$. Using the parameters defined earlier, we can estimate that the maximum coupling $g_{E_2}$ found in our system ($g_{E_2}/2\pi \approx 1.5$~MHz) can reach near-unity cooperativity $C_{E_2}=4g_{E_2}^2/(\gamma_{\text{th}} \Gamma_{x/y})\approx 0.7$, suggesting the system approaches the regime of coherent exchange of excitations between the cavity and the NV.

This coupling also offers a much more efficient pathway to implementing cooling of the acoustic resonator by tuning the mode energy to the energy splitting between the orbital states $\ket{x}_s$ and $\ket{y}_s$: $\omega_x-\omega_y\approx \Omega_\alpha$ \cite{rabl13prl}. First we optically populate the lower energy $\ket{y}_s$ state by optical driving at tuned to the  transition between ground and $\ket{y}_s$ states at $\omega_y$ with Rabi frequency $\Omega_R$. The NV then transitions to the higher-energy orbital $\ket{x}_s$ state by absorbing a cavity phonon, and subsequently relaxes emitting a photon at $\omega_x$, and cooling the system at rate $\Gamma_{E_2} = 4 g_{E_2}^2 \Omega_R^2/\Gamma_{x/y}^3$ \cite{rabl13prl}, to a final population given approximately by $n_\text{fin}=\gamma_\text{th}/\Gamma_{E_2}\approx 1.5$ for $g_{E_2}/2\pi=1.5$~MHz.

\subsection{Enhancing SiV spin-phonon coupling in other designs of subwavelength cavities}
\label{SiV}
A similar problem of engineering coherent coupling between an acoustic mode of a PnCW cavity and another widely analyzed colour defect in diamond --- the silicon vacancy (SiV) --- was investigated by Meesala \textit{et al.} in Ref.~[\citenum{PhysRevB.97.205444}]. SiVs are an attractive alternative to NVs for both information storage, and spin-orbit coupling. Thanks to the strong spin-orbit coupling, spin states within the ground electronic state manifold of SiV exhibit simultaneous lower dephasing rate (of about $\Gamma_{\text{SiV,deph}}/2\pi\approx 4$~MHz at 4~K and $100$~Hz at 100~mK \cite{PhysRevLett.119.223602}) and larger strain susceptibility (about 1.8~GHz) than NVs. Meesala \textit{et al.} noticed that the strain in a triangular crystal waveguide \cite{Burek:16} can be resonantly localized to a small volume near the surface of the diamond for a flapping mode of the waveguide \cite{Burek:16}. This localization supports resonant spin-phonon coupling with coupling $g_{\text{SiV}}/2\pi\approx 3$~MHz, and cooperativity of $C_{\text{SiV}}\sim 1$ for resonators with $Q\sim 10^5$ or $Q\sim 10^3$, at 4~K and 100~mK temperatures, respectively. This brief analysis indicates that the localization of the strain field found in structures developed by Burek \textit{et al.} \cite{Burek:16} yields effective mode volumes $\Veff\sim 10^{-3}\lambda_p^3$, which are comparable to those reported here, albeit achieved by a very different, resonant effect.

Finally, we note that by replacing NV with SiV in our cavities, and focusing on the resonant spin-phonon interaction, we could reach cooperativities $C_{\text{SiV}}\sim 4g_{\text{SiV}}^2/(\Gamma_{\text{SiV,deph}}\kappa_b)\approx 110$, where we have taken $g_{\text{SiV}}= 2g_{E_2}\approx 2\pi \times 3$~MHz (to reflect the larger strain susceptibility of SiV), for 4~K temperature and $\sim 8000$ for 100~mK.

\section{Fabrication considerations}
\label{section:fab}
Maximum values of the cooperativities depend critically on the acoustic decay rate, or quality factor of the acoustic cavity mode. To date, to the authors' best knowledge, the maximum $Q_m\sim 7000$ was reported by Burek \textit{et al}.~\cite{Burek:16} for few-GHz PnC cavities fabricated in bulk etched single-crystalline diamond. In a recent contribution by Cady \textit{et al.}~\cite{jayich2019qscitech}, authors reported on fabrication of diamond PnC cavities using the diamond-on-insulator technique with $Q_m\sim 100$, pointing to the significant losses induced by deviations and imperfections of the fabricated structures. Both of these reports cite quality factors measured at a room temperature, and should be further enhanced in cryogenic environment. Possible improvements could be achieved by embracing the recently developed concepts of soft-clamping \cite{tsaturyan2017ultracoherent} and strain engineering \cite{ghadimi2018elastic}. Furthermore, in materials for which fabrication techniques are more mature, such as silicon, much higher quality factors of PnC resonators up to $10^{10}$ were reported recently \cite{maccabe2019phononic}, suggesting that GHz acoustic vibrations can be used as quantum memories \cite{PhysRevLett.123.250501} rather than transducers \cite{PhysRevX.5.031031}. Finally, in a recent theoretical proposal, Neuman \textit{et al.} \cite{neuman2020phononic} proposed utilizing heterogeneous structures in which silicon phononic crystals would be interfaced with diamond patches hosting atomic defects.

\section{Conclusions}

In summary, we propose a simple design of an acoustic cavity capable of localizing GHz mechanical modes into ultrasmall volumes of about $10^{-4}\lambda_p^3$. Since these cavities are implemented as defects in quasi-one-dimensional phononic crystals, and the localization mechanism is non-resonant, the cavity frequencies can be readily tuned across the few-GHz range by changing geometric parameters. The quality factor is determined by the efficiency of suppression of transmission in the phononic Bragg mirrors.

We further find that such state-of-the-art cavities should, thanks to the significant spatial and spectral confinement of the acoustic mode, offer an attractive platform for implementing efficient coherent control over states of the atomic defects in diamond (NV or SiV) susceptible to the external strain. In particular, for the designs analyzed here, the resonant NV-phonon coupling operates in the high-cooperativity regime, opening a pathway to an efficient ground state resonant cooling of the cavity mode by a single NV. We also predict that similar setups could provide and even larger cooperativity of resonant coupling between phonons and the spin states of a SiV.

The proposed design of the cavity and strain localization mechanism can be further refined and implemented in more robust architectures, including cascaded acoustic cavities for indistinguishable phonon emission \cite{PhysRevLett.122.183602},  
quasi-two-dimensional phononic topological crystals \cite{PhysRevB.97.020102}, and acoustic buses for efficient transfer of a quantum state between distant emitters \cite{PhysRevLett.120.213603}. They should also be readily adapted to simultaneously co-localize high-Q optical mode \cite{weiss16acsphot,englund17prl,weiss18sciadv} to enable more efficient optical control of the defects.

\begin{acknowledgments}

M.K.S. would like to thank Ruben Esteban and Miros\l{}aw R. Schmidt for fruitful and insightful discussions. Authors acknowledge funding from Australian Research Council (ARC) (Discovery Project DP160101691) and the Macquarie University Research Fellowship Scheme (MQRF0001036). 

\end{acknowledgments}

\appendix

\section{Acoustic field operators and mode volume definition}
\label{Appendx.Quantization}
Both the displacement field $\hat{\mathbf{u}}_\alpha$ and strain tensor operators $\hat{\etensor}_\alpha$ corresponding to any given mode $\alpha$, have a representation given by the classical field $\mathbf{u}_\alpha$ and tensor $\etensor_\alpha$, respectively:
\begin{equation}
    \hat{\etensor}_\alpha(\mathbf{r}) = \sqrt{\frac{\hbar }{2\Omega_\alpha}}\left\{\etensor_\alpha(\mathbf{r}) b_\alpha + \left[\etensor_\alpha(\mathbf{r})\right]^* b_\alpha^\dag\right\}.
\end{equation}
\begin{equation}
    \hat{\mathbf{u}}_\alpha(\mathbf{r}) = \sqrt{\frac{\hbar }{2\Omega_\alpha}}\left\{\mathbf{u}_\alpha(\mathbf{r}) b_\alpha + \left[\mathbf{u}_\alpha(\mathbf{r})\right]^* b_\alpha^\dag\right\}.
\end{equation}
Then the total energy of the mode can be re-written as
\begin{align}
    \hat{H} &= \int_V \hat{h}(\mathbf{r}; \alpha) \text{d}\mathbf{r} \\ \nonumber 
    &= {\frac{\hbar}{2\Omega_\alpha}}\int_V\left\{ b_\alpha b_\alpha^\dag h(\mathbf{r}; \alpha)+b_\alpha^\dag b_\alpha h(\mathbf{r}; \alpha) \right\} \text{d}\mathbf{r} \\ \nonumber
    &= \left( 2b_\alpha^\dag b_\alpha +1\right){\frac{\hbar}{2\Omega_\alpha}}\int_V h(\mathbf{r}; \alpha) \text{d}\mathbf{r}  \\ \nonumber
    &= \hbar \Omega_\alpha \left( b_\alpha^\dag b_\alpha +\frac{1}{2}\right){\frac{1}{\Omega^2_\alpha}}\int_V h(\mathbf{r}; \alpha) \text{d}\mathbf{r} \\ \nonumber&= \hbar \Omega_\alpha \left( b_\alpha^\dag b_\alpha +\frac{1}{2}\right), \\ \nonumber
\end{align}
where in the last step we equated the derived Hamiltonian with the expected bosonic Hamiltonian, effectively defining the normalization of the elastic field/strain tensor. Since the energy is equally distributed between the kinetic and strain energy components, the normalization condition can be re-written by considering either form. For example, by focusing on the kinetic energy, the normalization is
\begin{equation}
    \frac{1}{\Omega^2_\alpha}\int_V h(\mathbf{r}; \alpha) \text{d}\mathbf{r}= \int_V \rho(\mathbf{r}) |\mathbf{u}_\alpha(\mathbf{r})|^2 \text{d}\mathbf{r} = 1.
\end{equation}
This is the explicit normalization used in our earlier work on elastic Purcell effect \cite{schmidt18prl} with interaction between an emitter modelled as a local harmonic force and the displacement field of a resonator.

In this contribution, it is convenient to express the normalization of the strain tensor by considering the contribution from strain energy, i.e.
\begin{equation}
    \frac{1}{\Omega^2_\alpha}\int_V h(\mathbf{r}; \alpha) \text{d}\mathbf{r}=\frac{1}{\Omega_\alpha^2}\int_V \left[\etensor_\alpha(\mathbf{r})\right]^*:\ctensor(\mathbf{r}):\etensor_\alpha(\mathbf{r}) \text{d}\mathbf{r} = 1.
\end{equation}
Looking back to Eq.~\eqref{Veff.def}, and noting that the maximum energy density in the resonator considered in this work is determined by the maximum of strain energy, we can approximate $\text{max}~|\etensor|\approx \sqrt{\Omega_\alpha^2/(|\ctensor|\Veff)}$, yielding a crude characteristic of the coupling coefficients $g_E\sim \Veff^{-1/2}$. This translates to a Purcell-like enhancement of the cooperativities (both for the resonant and the parametric couplings), inversely proportional to $\Veff$.

\section{Collective effects in cooling protocols}
\label{Appendix.CollectiveEffects}

Coupling to multiple ($N$) NVs provides a linear increase of the efficiency of the phonon cooling cooling $\Gamma_{E_1/E_2}^{\text{(coll)}}=N \Gamma_{E_1/E_2}$ for identical NVs. That is because, in the \textit{inverse Purcell effect} (defined by the hierarchy of decay rates $\Gamma_{x/y}\gg \gamma_\text{th}$) \cite{PhysRevB.94.014506}, the NVs behave as a collection of uncorrelated reservoirs for the cavity phonons. 

It would be thus tempting to enhance the cooling capabilities by simply increasing the number of NVs inside the cavity. Let us consider that idea, assuming that the NVs are positioned near the centre of the cavity with a constant density $\rho_\text{NV}$, and their individual couplings to the cavity mode can be approximately considered as identical. Under these assumptions, the number of NVs $N$ scales with \Veffa, and the collective cooling rates $\Gamma_{E_i}^{\text{(coll)}}$ scale with the number of photons $N \approx \rho_\text{NV} \Veffa$. However, since the single-NV cooling rates $\Gamma_{E_i}$ are inversely proportional to the effective mode volume through the localization mechanism ($\Veffa \propto 1/\sqrt{g_{E_i}} \propto 1/\Gamma_{E_i}$), the dependence of collective cooling rates on \Veffa~approximately cancels out. Therefore, the cooling mechanisms would not be considerably enhanced by increasing the dimensions of the cavity.

\section{Strain tensor transformation between laboratory and NV coordinate systems}
\label{Appendix.AxisRotation}

The strain tensor $\etensor^{(\text{lab})}$ of mode $\alpha$, calculated numerically and exported from COMSOL~\cite{comsol}, is given in the laboratory coordinate system $(X,Y,Z)$. For the calculation of coupling parameters in Section~\ref{sec:coupling} this must be transformed to the coordinate system of the NV $(x,y,z)$. That system is determined by selecting the $z$ axis along the vacancy - nitrogen direction, and choosing a vacancy-adjacent carbon atom that determines the $x$ axis (as a projection of the vacancy-carbon axis onto the plane perpendicular to $z$). Both $(X,Y,Z)$ and $(x,y,z)$ can be conveniently expressed in terms of crystallographic directions using local rotations from the crystallographic to laboratory coordinates $R_{\text{cryst}\rightarrow \text{lab}}$ and from the crystallographic to NV coordinates $R_{\text{cryst}\rightarrow \text{NV}}$, around the vacancy. For the particular choice of laboratory system discussed in the text, we have
\begin{equation}
    R_{\text{cryst}\rightarrow \text{lab}} = 
    \frac{1}{\sqrt{2}}\begin{bmatrix}
        1 & 1 & 0 \\
        -1 & 1 & 0 \\
        0 & 0 & \sqrt{2}
    \end{bmatrix}.
\end{equation}
Similarly, for a specific NV coordinate system with  $z$ and $x$ axes along $[\bar{1}\bar{1}\bar{1}]$ and $[{1}{1}\bar{2}]$, analyzed in Fig.~\ref{fig:FigCouplings}, we have
\begin{equation}
    R_{\text{cryst}\rightarrow \text{NV}} = 
    \frac{1}{\sqrt{6}}\begin{bmatrix}
         1 & 1 & -2 \\
        \sqrt{3} & -\sqrt{3} & 0 \\
        -\sqrt{2} & -\sqrt{2} & -\sqrt{2}
    \end{bmatrix}.
\end{equation}
Strain tensor $\etensor$ is transformed from the representation in the laboratory to the NV system of coordinates as
\begin{equation}
    \etensor^{(\text{NV})} = R_{\text{lab}\rightarrow \text{NV}}\cdot \etensor^{(\text{lab})}\cdot \left(R_{\text{lab}\rightarrow \text{NV}}\right)^{-1},
\end{equation}
with $R_{\text{lab}\rightarrow \text{NV}} = R_{\text{cryst}\rightarrow \text{NV}} \left(R_{\text{cryst}\rightarrow \text{lab}}\right)^{-1}$.

\section{Coupling for other orientations of the NV coordinate system}
\label{appendix.orientations}

\begin{figure}[htbp!]
\centering
\includegraphics[width=\linewidth]{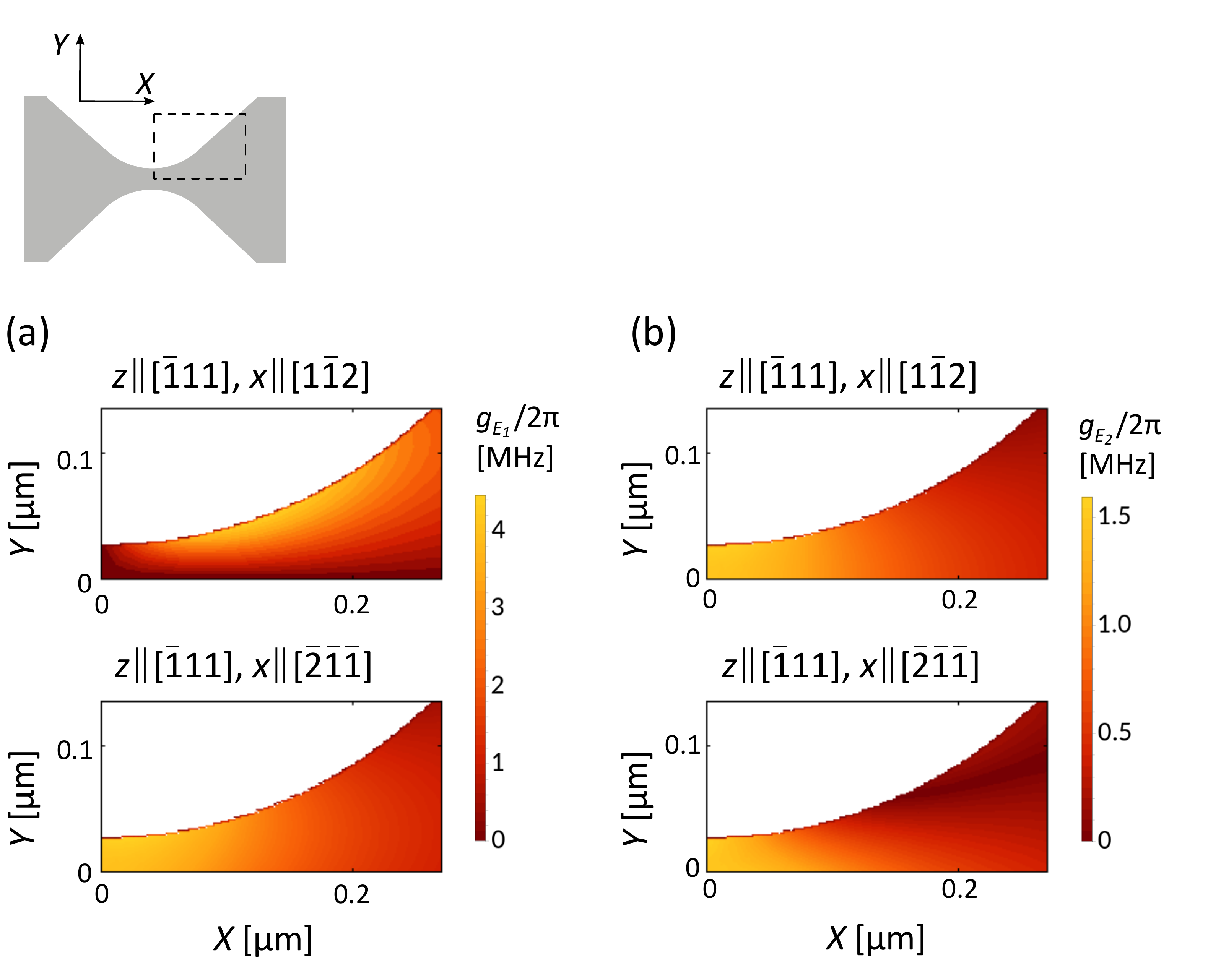}
 \caption{Maps of coupling parameters (a) $g_{E_1}$ and (b) $g_{E_2}$ for NVs positioned near the center of the cavity (see schematic at the top of the figure), and two different NV coordinate systems ($x$ axis) determined by the choice of a particular carbon atom. For all the cases we choose $z$ axis as parallel to the $[\bar{1}11]$ crystallographic orientation, in upper panels $x$ is parallel to $[1\bar{1}2]$ (as in the text), while in the lower panels --- to $[\bar{2}\bar{1}\bar{1}]$. As in Fig.~\ref{fig:FigCouplings}(b,c) NV are assumed to be positioned 5~nm below the surface of the diamond.}
\label{fig:FigCouplingsOtherC}
\end{figure}

\subsection{Coupling to static strain}

The Hamiltonian $H_\text{strain-NV}$ describes the effect of both the static strain --- either intrinsic to the structure, or applied to tune its mechanical response --- and dynamic, GHz strain of the acoustic cavity mode. Static strain shifts ($A$), splits ($E_1$) and couples ($E_2$) molecular orbitals $\{\ket{x},~\ket{y}\}$ of the NV, which are, in the absence of strain, defined by the local NV coordinate system $xyz$. This Hamiltonian in the basis $\{\ket{x},\ket{y}\}$, is given by 
\begin{equation}
    H_\text{strain-NV} = \begin{bmatrix}
        g_A+g_{E_1} & g_{E_2} \\
        g_{E_2} & g_A-g_{E_1}
    \end{bmatrix} \otimes (b_\alpha^\dag+b_\alpha),
\end{equation}
and can be then diagonalized, revealing strain-shifted energies, and a new basis of orbitals $\{\ket{x}_s,~\ket{y}_s\}$. This diagonalization naturally yields identical results irrespective of the initial orientation of the NV's $x$ axis.

\subsection{Coupling to dynamical strain}

This observation allows us to reconcile the effect of coupling to the dynamical strain field of the acoustic mode. While for a selected $z$ axis of the NV, the coupling coefficients $\{g_A,g_{E_1},g_{E_2}\}$ describing projection of the strain onto the irreducible representations of the $C_{3v}$ group, again differ with the choice of the NV $x$ axis, arbitrary states $\{\ket{x},~\ket{y}\}$ are mixtures of contributions from non-degenerate $\{\ket{x}_s,~\ket{y}_s\}$ orbitals, and need to be projected onto these orbitals to yield meaningful results.

This means that identifying $g_{E_1}$ calculated for an arbitrarily chosen $x$ axis with the general response of the strained NV is incorrect. Physically-relevant coupling parameters $g_{E_1,s}$ and $g_{E_2,s}$ quantify the interaction in the $\{\ket{x}_s, \ket{y}_s\}$ basis and coordinate system. Their values can be found in a similar way as for the static strain, by expressing the Hamiltonian $H_{E_1}+H_{E_2}$ in the basis of orbitals $\{\ket{x},\ket{y}\}$, and transforming it to the $x_sy_sz$ coordinate system. The diagonal and off-diagonal elements of that matrix will yield $g_{E_1,s}$ and $g_{E_2,s}$, respectively, and will be independent of the original choice of coordinate systems $xyz$.

This transformation will in general mix $E_1$ and $E_2$ couplings, and analyzing their effects separately, as we have done in the manuscript, is only justified under assumption that the physical coordinate system $x_sy_sz$ aligns with the selected local NV system $xyz$. 


We should also ask what happens in the absence of the static strain, when the degeneracy of the orbital states is not lifted. In this case, we can diagonalize the interaction Hamiltonian similarly as for the static case, with the basis $\{\ket{x}_s,\ket{y}_s\}$, in which the mixing $E_2$-like interactions vanish, and where as previously, the energies of eigenstates are independent of the local coordinate system of the NV. To the optical interrogation, the NV then behaves like two mutually decoupled two-level systems, parametrically coupled to the same mechanical mode of the cavity.

\bibliography{references}

\end{document}